\documentclass[hyper]{JHEP} 

\usepackage{epsfig}

\newcommand\fverb{\setbox\pippobox=\hbox\bgroup\verb}
\newcommand\fverbdo{\egroup\medskip\noindent%
			\fbox{\unhbox\pippobox}\ }
\newcommand\fverbit{\egroup\item[\fbox{\unhbox\pippobox}]}
\newbox\pippobox
\title{The Schrodinger Wave Functional 
and Closed String Rolling Tachyon}
\author{ J. Kluso\v{n}
\footnote{On leave from Masaryk University, Brno}\\
Institute of Theoretical Physics, University of Stockholm, SCFAB\\
SE- 106 91 Stockholm, Sweden \\
and \\
Institutionen f\"or teoretisk fysik\\
BOX 803, SE- 751 08 
Uppsala, Sweden \\
E-mail: \email{josef.kluson@teorfys.uu.se}}

\preprint{\hepth{0308023}}

\abstract{In this short note we apply Schrodinger
picture description of the minisuperspace
approach to the closed string tachyon
condensation. We will calculate the rate of
produced closed string  and we will 
show that the  density 
of high massive closed string modes
reaches the string
density  in time of order  one
in string units.}

\def\ket #1{\left|#1\right>}

\def\bk{\mathbf{k}}
\def\bx{\mathbf{x}}
\def\by{\mathbf{y}}
\def\bz{\mathbf{z}}

\def\ss{\sin \frac{\tau}{\sqrt{2}}}
\def\st{\sinh \frac{\tau}{\sqrt{2}}}
\def\st2{\sinh^2 \frac{\tau}{\sqrt{2}}}
\def\ss2{\sin^2 \frac{\tau}{\sqrt{2}}}

\def\ik{\frac{d\bk}{(2\pi)^{D}}}
\def\tg{\tilde{G}}

\def\hp{\hat{\pi}}
\def\hf{\hat{\phi}}
\begin{document}
\section{Introduction}\label{first}
A basic issue in string theories that have tachyons
in tree-level spectra is their fate.  The presence of
the tachyons leads to the condensation processes
in given theories. Most of the results are based
on the conjecture that the world-sheet models
that realize these initial and final states of
these condensation process are related through
renormalization group  (RG) flow on the world-sheet.
This relation between space-time evolution and
RG flow is still not well understood, especially in
case of closed tachyons (For review, see
\cite{Martinec:2002,Gutperle:2002,Moore:2003vf}.). A lot of 
open issues can only be studied through
the construction of exact  time-dependent backgrounds.
For the open string tachyon such an investigation
was invented by A. Sen \cite{Sen:2002an,Sen:2002in,Sen:2002nu}
and further developed in 
\cite{Gutperle:2002ai,Strominger:2002pc,Gutperle:2003xf,Maloney:2003ck}
(Recent works considering rolling tachyon in 
open string theory 
are \cite{Hashimoto:2003qx,Hashimoto:2002sk,Sen:2002qa,
Kutasov:2003er,Lambert:2003zr,McGreevy:2003kb,
Moeller:2003gg,Gaiotto:2003rm,Aref'eva:2003qu,
Okuda:2002yd,Mukhopadhyay:2002en,Moeller:2002vx,
Sen:2002vv,Chen:2002fp,
Minahan:2002if,Sugimoto:2002fp,
Fujita:2003ex,Rey:2003xs,Rey:2003zj,Klebanov:2003km,
McGreevy:2003ep,Constable:2003rc,Martinec:2003ka,
Gubser:2003vk,Takanayagi:2003,Sen:2003xs,
Douglas:2003up,Gaiotto:2003yf}.)

The condensation processes of closed tachyon are very difficult
in picture, mainly because their drastic effect they have on
space-time itself.  Some results have been obtained recently
in case of certain localized closed string tachyons
\cite{Adams:2000,Vafa:2001,Harvey:2001}.
On the other hand it was shown in
\cite{Strominger:2003ct} that there are exact, time-dependent classical
solutions tachyonic closed string theories describing homogeneous
tachyon condensation. This corresponding CFT on the worksheet
was named as timelier Liouville theory. It is described by the
action
\footnote{We have suppressed spatial direction and set 
$\alpha'=1$.}
\begin{equation}\label{S}
S=\frac{1}{4\pi}\int d^2\sigma
\left(-(\partial X^0)^2+4\pi\mu e^{2\beta X^0}\right) \ .
\end{equation}
This theory has negative norm boson and central charge
$c=1-6q^2 \ , q=\beta-\frac{1}{\beta}$. 
This corresponds to a real  dilaton with timelier slope
$q$  however  as in \cite{Strominger:2003ct} we will be mainly
interested in $q=0 \ , \beta=1$. The potential
term in (\ref{S}) can be interpreted as  closed string
tachyon field that grows exponentially in time. At early
times $X^0\rightarrow -\infty$ this term is small and we
have ordinary string action in flat space. 

The main problem with (\ref{S}) is that this action
is not positive definite and hence does not fully define
the associated functional integral in CFT. It is natural
to define timelier theory by analytic continuation
$\phi=-iX^0$ and $b=i\beta$ from the ordinary
Spacetime Liouville theory with positive definite worksheet
action
\begin{equation}
S=\frac{1}{4\pi}\int d^2\sigma
\left((\partial \phi)^2+4\pi \mu
e^{2 b\phi}\right) \ .
\end{equation} and background
charge
$Q=b+1/b$. The central charge is
$c=1+6Q^2$. This strategy has been used recently
in  \cite{Strominger:2003ct,
Schomerus:2003vv} where many intersting results
can be found. 

Since the worksheet action (\ref{S}) describes the
propagation of the string in the time-dependent background 
we can expect  the closed string
pair 
production. The rate of pair  production has been calculated in
\cite{Strominger:2003ct} in  minisuperspace approximation
where it was shown that it diverges exponentially. 
According to \cite{Strominger:2003ct} this result implies
that the gas of  produced  closed strings will reach
the string density and the string perturbation theory will 
break down in time of order string units. 
The calculation of the particle production  
 was performed using
the  canonical formalism that is well known from
the study of the  quantum field theory
in curved peacetime \cite{Birrell:ix,Fulling:nb,Wald:yp}
where quantum fields are considered as Heisenberg
operators that explicitly depend on time while
the states of the system do not evolve.
As companion to this approach we would like
to calculate the rate of the closed string pair production
in the Schr\"odinger representation
of the quantum field theory.
 The
reason why we perform this analysis is that
the Schr\"odinger  
picture  provides a simple and
intuitive description of vacuum states in quantum field
theory in situations where the background metric
is time-dependent or in case where some parameters
of quantum field theory depend on time. The 
vacuum states are explicitly  specified single,
 possibly time-dependent,
kernel function satisfying a differential equation with
prescribed boundary conditions. This makes no reference
to the assumed spectrum of excited states and so circumvents
the difficulties of the conventional canonical description
of a vacuum as a "no-particle" state with respect to the
creation and annihilation operators defined by a particular
mode decomposition of the field, an approach which is not
well suited to the time-dependent problems.

The main goal of this paper is to formulate
 the Schr\"odinger picture description
of the minisuperspace approximation to the closed string 
rolling tachyon. Using this powerful approach
to the quantum field theory in time-dependent
background we will 
obtain an expression for the density 
of particles produced during the rolling 
tachyon evolution.
Then we will estimate the time when the particle
density reaches the string scale and we will show that
this situation occurs very quickly after beginning of
the rolling tachyon process and leads to the 
breakdown of string perturbation theory. 

This paper is organized as follows. In the next section
(\ref{second})
we formulate the Schr\"odinger picture description
of the minisuperspace approach to the closed string
rolling tachyon background. Then we will calculate the rate
of particle production and in the asymptotic future
we find agreement with \cite{Strominger:2003ct}.
As a application of this formula we will estimate the
time when the density of produced closed string modes
reaches string density. In conclusion 
(\ref{third}) we outline
our results. 
\section{Minisuperspace approximation and 
Schr\"odinger picture description}\label{second}
In this section we will describe the closed string tachyon
condensation in the minisuperspace approximation
\cite{Strominger:2003ct}.
In the minisuperspace approximation we retain only the zero mode
$x^0$ of $X^0$. The on-shell condition for the closed string excitation is
Klein-Gordon  equation with time-dependent mass
\cite{Strominger:2003ct} 
\begin{equation}\label{minieq}
\left[\frac{\partial^2}{(\partial x^0)^2}+
k^2+2\pi\mu e^{2x^0}+2(N_L+N_R-2)\right]
\phi(x_0)=0 \ .
\end{equation}
Using 
$\phi(\bx,t)=\int \ik \phi(t)e^{i\bk\bx}$
  we can write (\ref{minieq}) as 
\footnote{Our convention is
$\eta_{\mu\nu}=\mathrm{diag}(-1,\dots, 1) \ ,
\mu\ , \nu=0,\dots, D$, $a,b,c,\dots=1,\dots,D$. We also
denote $x^0=t$ and $\bx=(x^1,\dots,x^D)$ and
$m^2=2(N_L+N_R-2)$, where $N_{L,R}$ are the
left and the right-moving oscillators contributions.}
\begin{equation}
\left[\partial_t^2-\delta^{ab}\partial_a\partial_b+
2\pi\mu e^{2t}+m^2\right]\phi(\bx,t)=0 \ .
\end{equation}
It is clear that this equation of motion 
can be obtained by variation of the following action
\begin{equation}\label{Ss}
S=-\int dt L=
-\frac{1}{2}\int dt d\bx
\left[\eta^{\mu\nu}\partial_{\mu}\phi
\partial_{\nu}\phi+m^2(t)\phi^2\right] \ ,
\end{equation}
where 
\begin{equation}
m^2(t)= 2\pi\mu e^{2t}+m^2 \ .
\end{equation}
The action (\ref{Ss}) describes the scalar field with time-dependent
mass. The quantisation of such field can be performed in 
the same way as in the case of  the quantum
field theory in curved space-time 
\cite{Birrell:ix,Fulling:nb,Wald:yp}, for recent 
analysis, see \cite{Strominger:2002pc,Gutperle:2003xf,
Maloney:2003ck,Strominger:2003ct}.
As companion to this approach 
 we apply Schr\"odinger picture description of the quantum
field theory defined by 
(\ref{Ss}). The similar approach
  was recently used for minisuperspace
description of S-branes in \cite{Kluson:2003sh}
\footnote{For nice review of Schr\"odinger picture
description of quantum field theory in curved
space-time, 
see \cite{Long:1996wf,Guven:1987bx,Eboli:1988qi}.}.

We begin with   the  action (\ref{Ss}) 
from which we obtain the canonical momentum conjugate
to $\phi(t,\bx)$ 

\begin{equation}
\pi(t,\bx)=\frac{\delta L}{\delta \dot{\phi}(\bx,t)}
=\dot{\phi}(t,\bx) \ 
\end{equation}
and the Hamiltonian
\begin{equation}
H=\int d\bx\left(\pi \dot{\phi}-L\right)=
\frac{1}{2}\int d\bx
\left(\pi^2+\eta^{ab}\partial_{a}\phi
\partial_b\phi+m^2(t)\phi^2\right) \ .
\end{equation}
The system can be  quantised by treating the
fields as operators and imposing appropriate commutation
relations. This involves choice of a foliation
of a space-time in a succession of
space-like hypersurfaces. We choose these to
be the hypersurfaces of fixed $t$ and
impose equal-time commutation relations 
\begin{equation}
[\hf(\bx,t),\hp(\by,t)]=
i\delta (\bx-\by) \ , [\hf(\bx,t),\hf(\by,t)]=
[\hp(\bx,t),\hp(\by,t)]=0 \ .
\end{equation}
In the Schr\"odinger picture we take the basis vector of the state 
vector space to be the eigenstate of the field operator $\hf(t,\bx)$
on a fixed $t$ hypersurface, with eigenvalues $\phi(\bx)$
\begin{equation}
\hf(t,\bx)\ket{\phi(\bx),t}=
\phi(\bx)\ket{\phi(\bx),t} \ .
\end{equation}
Notice that the set of field eigenvalues $\phi(\bx)$
is independent of the value of $t$ labeling the
hypersurface.  In this picture, the quantum states
are explicit functions of time and are represented
by wave functionals $\Psi[\phi(\bx),t]$. Operators
$\hat{\mathcal{O}}(\hp,\hf)$ acting on these
states may be represented by
\begin{equation}
\hat{\mathcal{O}}(\hp(\bx),\hf(\bx))=
\mathcal{O}\left(-i\frac{\delta }{\delta
\phi(\bx)},  \phi(\bx)\right) \ .
\end{equation}
The Schr\"odinger equation which governs the evolution
of the wave functional  is
\begin{eqnarray}\label{schrod}
i\frac{\partial \Psi[\phi,t]}
{\partial t}=H\left(-i\frac{\delta }{\delta
\phi(\bx)},  \phi(\bx)\right)\Psi[\phi,t]=\nonumber \\
=\frac{1}{2}\int
d\bx \left[-\frac{\delta^2}{\delta \phi^2}
+\eta^{ab}\partial_a\phi\partial_b\phi+
m^2(t)\phi^2\right]\Psi[\phi,t] \ . 
\nonumber \\
\end{eqnarray}
To solve this equation, we make the ansatz, that 
 the vacuum functional is simple
Gaussian.  We therefore write
\begin{equation}\label{vacans}
\Psi_0[\phi,t]=
N_0(t)\exp\left\{-\frac{1}{2}\int
d\bx d\by \phi(\bx)G(\bx,\by,t)\phi(\by)\right\}=
N_0(t)\psi_0(\phi,t) \ ,
\end{equation}
where $N_0(t) , G(\bx,\by,t)$ obey following
equations
\begin{eqnarray}\label{kernel}
i\frac{\partial N_0(t)}{\partial t}=
\frac{N_0(t)}{2}\int d\bz
G(\bz,\bz,t) \ , \nonumber \\
i\frac{\partial G(\bx,\by,t)}
{\partial t}=
\int d\bz  G(\bz,\bx,t)G(\by,\bz,t)
-\left(\eta^{ab}\partial_a \partial_b
+m^2(t)\right)\delta(\bx,\by) \ .
\nonumber \\
\end{eqnarray}
Because  
the spatial sections are flat it is natural
to perform a Fourier transformation on
the space dependence of the field configuration
\begin{equation}
\phi(\bx)=\int \ik e^{i\bk \bx}\alpha(\bk) \ ,
\end{equation}
where reality of $\phi$ implies $\alpha^*(\bk)=
\alpha(-\bk)$. 
Similarly we can define $\delta/\delta \alpha(\bk)$
as 
\begin{equation}
\frac{\delta }{\delta \phi(\bx)}=
\int \ik e^{i\bk \bx}\frac{\delta }{
\delta \alpha(\bk)} \ ,
\end{equation}
where 
\begin{equation}
\frac{\delta \alpha(\bk)}{
\delta \alpha(\bk')}=
(2\pi)^{D}\delta
(\bk+\bk') \ .
\end{equation}
In $k$ space, the Hamiltonian is
\begin{equation}\label{hamk}
H=\frac{1}{2}\int
\ik \left[-\frac{\delta^2}{\delta \alpha(\bk)
\alpha(-\bk)}+
\Omega^2_{\bk}(t)\alpha(\bk)\alpha(-\bk)\right] 
\ ,
\end{equation}
where
\begin{equation}
\Omega^2_{\bk}(t)=m^2(t)+\omega^2_{\bk} \ ,
\omega^2_{\bk}=k^2+m^2 \ .
\end{equation}
For each $\bk$ the integrant
in (\ref{hamk}) represents
a harmonic oscillator with the 
time-dependent frequency $\Omega^2_{\bk}(t)$. 
After performing the Fourier transformation
for the kernel $G(\bx,\by,t)$
\begin{equation}
G(\bx,\by,t)=\int \ik
e^{i\bk(\bx-\by)}\tg(\bk,t) \ 
\end{equation}
the   kernel equation (\ref{kernel}) 
reduces to 
\begin{equation}
i\frac{\partial \tg(\bk,t)}
{\partial t}=\tg^2(\bk,t)-
\Omega^2_{\bk}(t) \ .
\end{equation}
This equation can be solved with the
ansatz
\begin{equation}
\tg(\bk,t)=-i\frac{\dot{\psi}_{\bk}(t)}{
\psi_{\bk}(t)} \ ,
\end{equation}
where  $\psi_{\bk}(t)$ obeys
\begin{equation}\label{psieq}
\ddot{\psi}_{\bk}+\Omega^2_{\bk}(t)
\psi_{\bk}=0  \ .
\end{equation}
Now the vacuum state functional  has
the form
\begin{equation}
\Psi[\phi,t]=N_0(t)
\exp \left(\frac{i}{2}\int \ik 
\alpha(-\bk)\frac{\dot{\psi}_{\bk}}{\psi_{\bk}} 
\alpha(\bk)\right) \ ,
\end{equation}
where 
\begin{equation}
N_0(t)=e^{-i\int^t dt'E_0(t')} \ ,
E_0(t)=\frac{1}{2}
V \int \ik \tg(\bk,t) \ .
\end{equation}
In the previous expression 
$V$ is spatial volume of D+1 dimensional
 space-time. 
For closed string rolling tachyon
the equation
(\ref{psieq}) has the form
\begin{equation}
\left[\partial^2_t+2\pi\mu e^{2t}+
\omega^2_{\bk}\right]\psi_{\bk}(t)=0
\ .
\end{equation}
The solution of the previous equation is
\begin{equation}
\psi^{in}_{\bk}=(\frac{\pi\mu}{2})^{i\omega_{\bk}/2}
\frac{\Gamma(1-i\omega_{\bk})}{\sqrt{2\omega_{\bk}}}
J_{-i\omega_{\bk}}(\sqrt{2\pi\mu}e^{t}) \ .
\end{equation}
For next purposes we also write the
second solution of the previous equation
\begin{equation}\label{out}
\psi^{out}_{\bk}(t)=\sqrt{\frac{\pi}{2}}
(ie^{\pi\omega_{\bk}})^{-1/2}H^{(2)}_{-i\omega_{\bk}}
(\sqrt{2\pi\mu}e^t)  \ . 
\end{equation}
The relation between these two modes is
\begin{equation}
\psi^{out}_{\bk}=A\psi^{in}_{\bk}+
B\psi^{in *}_{\bk} \ ,
\end{equation}
where
\begin{equation}
A=e^{\pi\omega_{\bk}+\pi i/2}B^*=
\sqrt{\frac{\omega_{\bk}\pi}{2}}
e^{\pi\omega_{\bk}/2
-i\pi/4}\left(
\frac{(\pi\mu/2)^{-i\omega_{\bk}/2}}
{\sinh \pi\omega_{\bk}
\Gamma(1-i\omega_{\bk})}\right) \ .
\end{equation}
Since for $t\rightarrow -\infty$ the 
mass term approaches standard flat space 
expression it is natural to demand that 
the vacuum state functional  
$\Psi[\phi,t]$ approaches the usual positive frequency 
Minkowski vacuum state 
\begin{equation}
\psi_0(\phi,-\infty)=
\exp\left(-\frac{1}{2}\int \ik
|\phi(\bk)|^2\omega_{\bk}\right) \ .
\end{equation}
This boundary condition for $\Psi[\phi,t]$
implies  following form of the kernel
\begin{equation}
\tg(\bk,t)=-i\frac{\dot{\psi}^{in*}_{\bk}(t)}
{\psi_{\bk}^{in*}(t)} \ .
\end{equation}
To see this note that for
 $t\rightarrow -\infty$ we have
\begin{equation}
\lim_{t\rightarrow -\infty}
\psi_{\bk}^{in*}(t)=
\frac{1}{\sqrt{2\omega_{\bk}}}e^{i\omega_{\bk}
t}  \ ,
\lim_{t\rightarrow -\infty}
\tg(\bk,t)=\omega_{\bk} \ .
\end{equation}
As in  \cite{Strominger:2003ct}
we would like to determine the rate  of the
particle production in the rolling
tachyon background. In the same way
as in \cite{Kluson:2003sh}
we introduce 
an operator of number of  particles with
momentum $\bk$  
\begin{eqnarray}
N(\bk,t)=
\frac{1}{2\Omega_{\bk}(t)}\left[
-\frac{\delta^2}{\delta \alpha(-\bk)
\alpha(\bk)}+\Omega^2_{\bk}(t)\alpha(\bk)
\alpha(-\bk)-\Omega_{\bk}(t)(2\pi)^D
\delta_{\bk}(0)\right] \ . \nonumber \\
\end{eqnarray}
To support the claim that $N(\bk,t)$ is
natural operator of the number of particles
with momentum $\bk$ at time $t$ note
 that  the Hamiltonian can be written as 
\begin{equation}\label{hata}
H(t)=
\int \ik \Omega_{\bk}(t)\left[
N(\bk,t)+\frac{V_{D}}{2}\right]
\ , V_{D}=(2\pi)^{D} \delta_{\bk}(0) \ 
\end{equation}
which has the form of the collection of Hamiltonians
of harmonic oscillators with the time-dependent 
frequency $\Omega_{\bk}(t)$ 
where the operator $N(\bk,t)$ 
counts the number of excited modes.

We would like to calculate 
the vacuum expectation value of 
$\left<N(\bk,t)\right>$. 
The calculation of $\left<N(\bk,t)\right>$
 is completely the same as
in case of S-brane dynamics 
\cite{Kluson:2003sh} where 
it was shown that
\begin{eqnarray}\label{nvac}
\left<N(\bk,t)\right>=(2\pi)^{D}\delta_{\bk}(0)
\frac{\left(\Omega_{\bk}(t)-\tg(\bk,t)\right)
\left(\Omega_{\bk}(t)-\tg^*(\bk,t)\right)
}{2\Omega_{\bk}(t)(\tg(\bk,t)+\tg^*(\bk,t))} \ .
 \nonumber \\
\end{eqnarray}
We can also define the vacuum expectation
value of the  spatial
density of the number of particles   as
\begin{equation}\label{Nden}
\left<\mathcal{N}(\bk,(t)\right>\equiv
\frac{\left<N_{\bk}(t)\right>}{V}=
\frac{\left(\Omega_{\bk}(t)-\tg(\bk,t)\right)
\left(\Omega_{\bk}(t)-\tg^*(\bk,t)\right)
}{2\Omega_{\bk}(t)(\tg(\bk,t)+\tg^*(\bk,t))} \ .
\end{equation}
Let us calculate $\left<\mathcal{N}(\bk,t)\right>$ 
for  the vacuum state functional
$\Psi^{in}[\phi,t]$ in the limit $t\rightarrow \infty$.
In the completely the same way as in
\cite{Kluson:2003sh} we obtain
that the  density of the number of
particle created  with momentum $\bk$
is equal to 
\begin{equation}
\left<\mathcal{N}(\bk,t)\right>=
|B_{\bk}|^2 \ .
\end{equation}
Now using the fact that
\begin{eqnarray}
B_{\bk}^*=\sqrt{\frac{\omega_{\bk}\pi}{2}}
e^{-\pi \omega_{\bk}/2}e^{-\frac{3\pi i}{4}}
\frac{(\frac{\pi\mu}{2})^{-i\omega_{\bk}/2}}{
\sinh \pi\omega_{\bk}
\Gamma(1-i\omega_{\bk})} \ , 
\nonumber \\
|\Gamma(1+i\omega_{\bk})|^2=
\frac{\pi \omega_{\bk}}{\sinh(\pi \omega_{\bk})} \ , 
\nonumber \\
|B_{\bk}|^2=\frac{e^{-\pi\omega_{\bk}}}{
2\sinh 2\pi \omega_{\bk}}
=\frac{1}{e^{\frac{\omega_{\bk}}{T}}-1}  \ ,
T=\frac{1}{2 \pi} \ 
\nonumber \\
\end{eqnarray}
we get the result that even if $\Psi^{in}[\phi,t]$ is pure
state  the number of particle produced at far future
is the same as in the thermal state of temperature $T=\frac{1}{2\pi}$.
Then  the vacuum
expectation value of the Hamiltonian 
\footnote{We omit  the zero-point energy contribution.}
 at far
future is
\begin{equation}
\left<H(t)\right>=
\int \ik\Omega_{\bk}(t)\left<N(\bk,t)\right>
=
V_D\sqrt{2\pi\mu}
\int \ik e^{\frac{2\pi t}{T_H}}
\frac{1}{e^{\frac{\omega_{\bk}}{T}}-1} \ ,
\end{equation}
where $T_H=\frac{1}{4\pi}$ is Hadegorn temperature.
However we must stress that the
 previous expression is not quiet correct since
we did not consider  a  degeneracy of
of closed string modes with initial energy
$\omega_{\bk}$. If we denote the
  degeneracy of closed string modes with initial 
energy 
 $\omega_{\bk}$ as
 $\rho(\omega_{\bk})$
then the total
energy density that is dumped in the string pairs
created in the rolling tachyon background 
is equal to
\begin{equation}
E_{tot}=\sqrt{2\pi\mu}
\int \ik \rho(\omega_{\bk})
e^{\frac{2\pi t}{T_H}}
\frac{1}{e^{\frac{\omega_{\bk}}{T}}-1} \ .
\end{equation}
For large $\omega_{\bk}$ we have
$\rho(\omega_{\bk})\sim e^{\frac{\omega_{\bk}}{T_H}}$ and
consequently 
\begin{equation}
E_{tot}\sim \int d\omega_{\bk}\omega_{\bk}^{D-1}
e^{\frac{2\pi t}{T_H}}
e^{\frac{\omega_{\bk}}{T_H}-\frac{\omega_{\bk}}{T}} \ .
\end{equation}
We immediately see that this integral 
diverges exponentially. As was firstly mentioned in
\cite{Strominger:2003ct} this divergence is much
more stronger then in case of S-branes where the
exponentials canceled and the divergence is
at most by power law. This result implies that
the gas of pair of produced closed strings 
will reach the string density in a time of one in string
units. To see this let us again consider the
vacuum expectation value of particle density
$\left<\mathcal{N}(\bk,t)\right>$ given in
(\ref{Nden}). As opposite to the previous case when
we calculated the rate of produced closed strings
at asymptotic future we will study this expression
in the far past when we can write
\begin{equation}
\psi^{in*}_{\bk}(t)\sim\frac{e^{i\omega_{\bk}t}}
{\sqrt{2\omega_{\bk}}}\left(1-\frac{\pi\mu
e^{2t}}{2(1+i\omega_{\bk})}\right)
\end{equation}
and consequently 
\begin{equation}
\tg(\bk,t)=-i\frac{\dot{\psi}^{in*}_{\bk}(t)}
{\psi^{in*}_{\bk}(t)}=
\omega_{\bk}+\frac{i\pi\mu e^{2t}}
{(1+i\omega_{\bk})} \  , 
\Omega_{\bk}(t)=\omega_{\bk}+
\frac{\pi\mu e^{2t}}{\omega_{\bk}} \ .  
\end{equation}
Using these results we can estimate
(\ref{nvac}) at far past as
\begin{equation}
\left<\mathcal{N}(\bk,t)\right>\sim
C_{\bk}
e^{4t} \  , 
\end{equation}
where $C_{\bk}$ is some numerical factor containing
powers of $\omega_{\bk}$. 
Since the density of states for high-energy modes
is $\rho\sim e^{\omega_{\bk}/T_H}$ we obtain
an  estimate for the  density of high-energy  particles
at far past as
\begin{equation}
\left<\mathcal{N}_{tot}(\bk,t)
\right>\sim \rho(\omega_{\bk})
\left<\mathcal{N}(\bk,t)\right>
\sim e^{\frac{\omega_{\bk}}{T_H}+
4t} \ .
\end{equation}
We see that this expression will reach
the string density $\left<\mathcal{N}_{tot}(\bk,t)
\right>\sim 1$  for
\begin{equation}
\frac{\omega_{\bk}}{T_H}+
4t\sim 0 \Rightarrow
t\sim -\frac{\omega_{\bk}}{T_H} \ . 
\end{equation}
The previous result implies that for
high-energy modes the time when the density
of produced closed string pairs reaches
the string density is very close to the beginning
of the rolling tachyon process roughly of
order one in string units. 
\section{Conclusion}\label{third}
In this short note we have formulated
the  Schr\"odinger picture description
of the quantum field theory that arises from 
 the minisuperspace 
approach to the closed string rolling tachyon 
which was considered previously in 
\cite{Strominger:2003ct,Schomerus:2003vv}.
We have also calculated the rate of
produced closed strings in rolling tachyon background
using Schr\"odinger representation of quantum
field theory and we have found completely
agreement with  \cite{Strominger:2003ct}.

To conclude this paper we would like to stress
that even if this paper brings  the modest contribution to
the research of the tachyon condensation and time-dependent
processes in string theory we hope that 
Schr\"odinger formulation of quantum field theory 
could be very useful  for addressing
these questions.  
 
{\bf Acknowledgment}
I would like to thank Prof. Ulf Lindstr\"om and
Prof. Ulf Danielsson for their support in my work. 
This work is partly supported by EU contract
HPRN-CT-2000-00122.

\end{document}